\documentclass{svproc}
\usepackage{url}

\usepackage{subcaption}
\usepackage{slashed}
\usepackage{amsmath}
\usepackage{amssymb,latexsym}
\usepackage{amsmath} 
\usepackage{graphicx}

\begin{document}
\mainmatter              
\title{Equivalence of two component spinor mechanism and four component spinor mechanism in top quark pair production}
\titlerunning{Equivalence of two component spinor mechanism and four component spinor mechanism }  
%
\author{Malvika Deo\inst{1} \and Anuradha Misra\inst{2}
Sharada Subramanian\inst{3} \and Radhika Vinze\inst{4}}
\authorrunning{Radhika Vinze et al.} 
%
\tocauthor{Malvika Deo, Anuradha Misra, Sharada Subramanian and Radhika Vinze}
\institute{Department of Physics, University of Mumbai, Mumbai, India\\
	\and
	UM-DAE Centre for Excellence in Basic Sciences (CEBS), Vidyanagari, Mumbai-400098, India \\
		\and
	Adyar, Chennai, India \\
	\and
Indian Institute of Science Education and Research Mohali, Knowledge City, Sector 81, SAS Nagar, Manauli, Punjab 140306, India  \\
\email{radhikavinze@iisermohali.ac.in},}

\maketitle              

\begin{abstract}
We calculate the $S$-matrix elements for the process $e^{+} e^{-}\rightarrow t \bar{t}$ mediated by SM photon, $Z$ boson and an additional $Z^{'}$ boson indicating the contribution from new physics. We calculate the amplitude square using two component spinor formalism and four component spinor formalism and show the equivalance of the results using the two formalisms. We also establish the relations between the couplings of $Z^{'}$ boson to fermions in the two component spinor formalism and in the four component spinor formalism.
\end{abstract}
\section{Introduction}
\hspace{-0.2cm}
One of the proposals to go beyond the Standard Model (SM) of particle physics is to extend SM by an additional $U(1)$ symmetry \cite{Langacker_2009}. The effects of $Z^{'}$ coupling the fermions in the SM can account for the discrepancy in the theoretically predicted and observed values of certain parameters \cite{Rizzo:2006nw}. 

The top quark plays a vital role in improving the understanding of electroweak sector of the SM. Looking at the uncertainty in the top quark properties such as top quark mass\cite{ATLAS:2018fwq}, top quark pair production cross section \cite{CMS:2018fks}, its coupling with other quarks etc can provide evidence for physics beyond the SM. Electron positron collider provides a clean environment as compared to hadron colliders.

At very high energies, as the initial and final state particles become effectively massless, the chirality and helicity of the particle become the same. Thus, one can shift to the helicity basis for writing the amplitude for the scattering process at very high energies. The helicity amplitude method is useful when there are large number of particles in the final state. In this formalism, the amplitude for the scattering process is written taking into account the left/right-handedness i.e. helicity of the particle.

Here, we present calculation of the $S$ matrix elements for the process $e^{+} e^{-} \rightarrow t \Bar{t}$ at leading order in an extension of SM which has an additional massive $Z^{'}$ boson in addition to the standard photon ($\gamma$) and a massive $Z$ boson. We have performed the calculation using two component formalism as well as the conventional four component formalism. Implementing a model independent approach for $Z^{'}$, we infer how the vector and axial vector couplings of the $Z^{'}$ boson in four component spinor mechanism are related to those in two component mechanism. 
	\vspace{-0.9cm}
\section{Calculation using helicity spinor formalism}
	\vspace{-0.1cm}
We consider scattering $e^{-} e^{+} \rightarrow t \Bar{t}$ mediated by SM photon ($\gamma$), $Z$ and NP $Z^{'}$. The Feynman diagrams are shown in Fig.\ref{fig:twocomp}. 
\begin{center} 
	\begin{figure}
	\vspace{-0.7cm}
\hspace{2.5cm}	\includegraphics[width=0.5\textwidth]{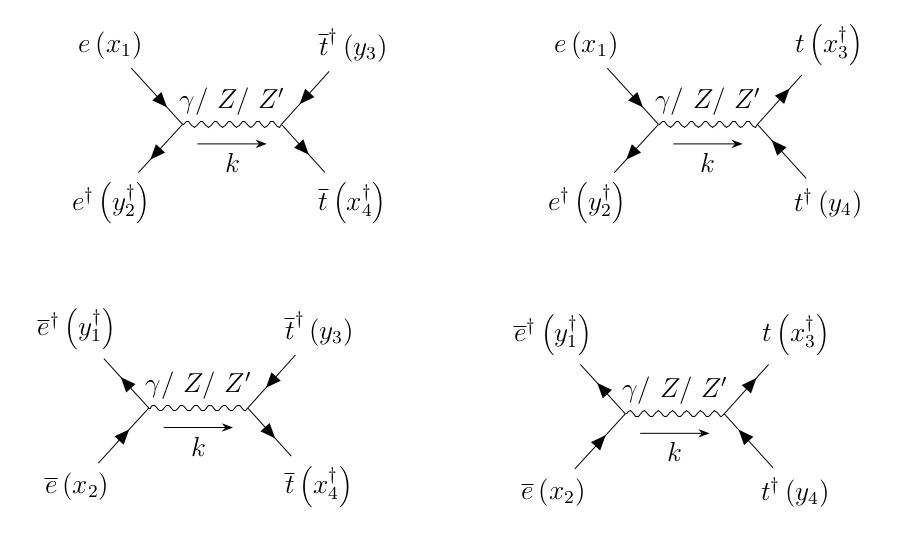} 
		\caption{Feynman diagram for Top quark pair production in $e^{+}\;e^{-}$ scattering in helicity spinor formalism. }
		\label{fig:twocomp}
	\end{figure}
	\vspace{-1.5cm}
\end{center} 

The amplitude for $e^{-}\; e^{+} \rightarrow t \;\bar{t}$ mediated by photon ($\gamma$), $Z$ boson, $Z^{'}$ boson exchange in two component spinor mechanism are ($Q_{e}=-1$, $Q_{t}= \frac{2}{3}$)
\begin{eqnarray}
	i\mathcal{M}_{\gamma} &=& \frac{-ig^{\mu\nu}}{k^2} \left[\left(iea_e\right)x_1\sigma_{\mu}y^{\dagger}_2 \left(-ie b_t\right)y_{3}\sigma_{\nu} x^{\dagger}_{4} 	+ \left(ie a_e\right) x_1\sigma_{\mu}y^{\dagger}_2 \left(-ie a_t\right)x^{\dagger}_{3}\bar\sigma_{\nu}y_{4} \right.\nonumber\\
	&~& \left. +\left(ie b_e\right)y^{\dagger}_{1}\bar\sigma_{\mu}x_{2}\left(-ie b_t\right)y_{3}\sigma_{\nu} x^{\dagger}_{4} + \left(ie b_e\right) y^{\dagger}_{1}\bar\sigma_{\mu}x_{2}\left(-iq a_t \right) x^{\dagger}_{3}\bar\sigma_{\nu}y_{4}\right] 
	\label{MG-two} \nonumber \\
	i\mathcal{M}_{Z} &=&  \frac{i e^2}{ s_W^2 c_W^2 } \left(\frac{g_{\mu\nu} - \frac{k_{\mu} k_{\nu}}{M_{Z'}^2}} {k^2-M_{Z}^2}\right) \nonumber \\
	&~&\hspace{-0.7cm}\left[\;a_e b_t x_1\sigma_{\mu}y^{\dagger}_2 y_{3}\sigma_{\nu} x^{\dagger}_{4}  - a_e a_t x_1\sigma_{\mu}y^{\dagger}_2 x^{\dagger}_{3}\bar\sigma_{\nu}y_{4} 	 + b_e b_t y^{\dagger}_{1}\bar\sigma_{\mu}x_{2}y_{3}\sigma_{\nu} x^{\dagger}_{4}
	- b_e a_t y^{\dagger}_{1}\bar\sigma_{\mu}x_{2} x^{\dagger}_{3}\bar\sigma_{\nu}y_{4}\right]  \nonumber
	\label{MZ-two}
\end{eqnarray}
\begin{eqnarray}
	i\mathcal{M}_{Z'} &=& i \frac{ \; \eta_1 \eta_2 e^2}{ s_W^2 c_W^2 } \left(\frac{g_{\mu\nu} - \frac{k_{\mu} k_{\nu}}{M_{Z'}^2}} {k^2-M_{Z'}^2}\right) \nonumber \\
	&~&\hspace{-0.7cm}\left[a^{'}_e b^{'}_t x_1\sigma_{\mu}y^{\dagger}_2 y_{3}\sigma_{\nu} x^{\dagger}_{4}
	- a^{'}_e a^{'}_t x_1\sigma_{\mu}y^{\dagger}_2 x^{\dagger}_{3}\bar\sigma_{\nu}y_{4}  + b^{'}_e b^{'}_t y^{\dagger}_{1}\bar\sigma_{\mu}x_{2}y_{3}\sigma_{\nu} x^{\dagger}_{4} - b^{'}_e a^{'}_t y^{\dagger}_{1}\bar\sigma_{\mu}x_{2} x^{\dagger}_{3}\bar\sigma_{\nu}y_{4}\right] \nonumber
	\label{MZp-two}
\end{eqnarray}
where we have used the conventions in \cite{Dreiner_2010}.
Here $a_i, b_i, a_i^{'}, b_i^{'}~;~ i = e,t~$ are the electron and top quark vector and axial vector couplings with $Z$, $Z^{'}$ respectively given in Table \ref{Table1}.
\begin{table}
	\centering
	\vspace{-0.4cm}
	\begin{tabular}{l c c c c} 
		\hline
		Mediator & $a_e$  & $b_e$ & $a_t$ & $b_t$\\ 
		\hline \vspace{-0.3cm} \\
		$\gamma$  & $-1$ & $-1$ & $\frac{2}{3}$ & $\frac{2}{3}$ \vspace{0.1cm} \\ 
		$Z$ & $-\frac{1}{2} + s_{W}^{2}$ & $s_{W}^{2}$ & $\frac{1}{2}-\frac{2 s_{W}^{2}}{3}$	& $\frac{2 s_{W}^{2}}{3}$ \vspace{0.1cm} \\  
		$Z^{'}$ & $a_e^{'}$  & $b_e^{'}$ & $a_t^{'}$ & $b_t^{'}$  \\ 
		\hline
	\end{tabular}
	\caption{Couplings of $Z$, $Z^{'}$ with electron and top quark in two component spinor mechanism}
	\label{Table1}
\end{table} 
\vspace{-1.0cm}	
\section{Calculation in four component spinor mechanism}
The Feynman diagram in conventional four component formalism is given in Fig.\ref{fig:fourcomp}
\begin{center} 
	\vspace{1.0cm}	
	\begin{figure}[h!]
		\centering
		\includegraphics[trim=3.0cm 21cm 2.5cm 9cm, width=1.0\linewidth]{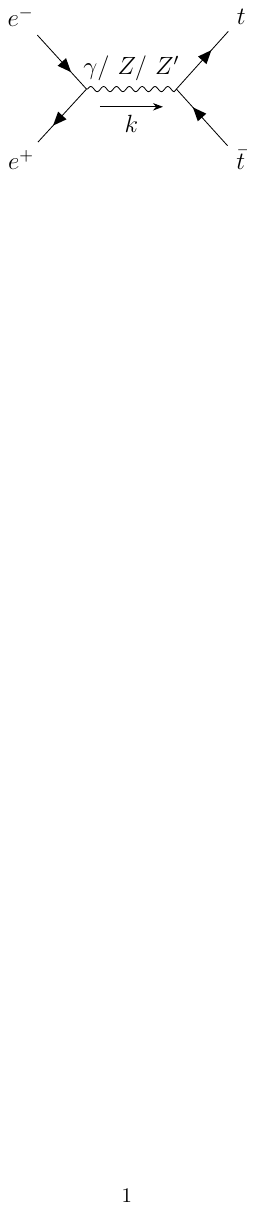} 
		\vspace*{-0.1in}  
		\caption{Feynman diagram for top quark pair production in $e^{+}\;e^{-}$ collision in four component spinor formalism}
		\label{fig:fourcomp}
		\vspace{-1cm}
	\end{figure}
\end{center} 
The amplitude for $e^{-}~e^{+} \rightarrow t~\bar{t}$ mediated by $\gamma$ in four component spinor mechanism mediated by photon ($\gamma$), $Z$ boson can be written using standard four component spinor conventions. The amplitude for $e^{-}~e^{+} \rightarrow t~\bar{t}$ mediated by $Z^{'}$ boson exchange with $\sin{\theta_{W}} = s_W,  \cos{\theta_{W}} = c_W $ is
\begin{eqnarray}
	\hspace{-0.5cm}i\mathcal{M}_{Z^{'}} &=&  \overline v(p_2) \left[\frac{-ie\eta_1}{2s_W c_W}\gamma^{\mu}\left(C_{V}^{'e}-C_{A}^{'e} \gamma^5 \right)\right] u(p_1)  \left(-i\frac{g_{\mu\nu} - \frac{k_{\mu} k_{\nu}}{M^2_{Z'}}}{k^2-M^2_{Z'}}\right) \overline u(p_3) \left[\frac{-ie\eta_2}{2s_W c_W}\gamma^{\nu}\left(C_{V}^{'t}-C_{A}^{'t} \gamma^5 \right)\right] v(p_4)  \nonumber
	\label{MZp-four}
\end{eqnarray}
where $C_{V}, C_{A}, C_{V}^{'}, C_{A}^{'}$ are the vector and axial vector couplings for $Z$ boson and $Z^{'}$ boson respectively with electron and top quark. These are listed in Table \ref{Table2}. 
\begin{table}
	\begin{tabular}{l c c c c} 
		\hline
		Mediator & $C_{V}^e$  & $C_{A}^e$ & $C_{V}^t$ & $C_{A}^t$\\ 
		\hline \vspace{-0.3cm} \\
		$Z$ & $2 s_{W}^2-\frac{1}{2}$ & $ -\frac{1}{2}$ & $\frac{1}{2}-\frac{4}{3} s_{W}^2$	& $\frac{1}{2}$ \vspace{0.1cm} \\  
		$Z^{'}$ & $C_{Ve}^{'}$  & $C_{A}^{e'}$ & $C_{V}^{t'}$ & $C_{A}^{t'}$  \\ 
		\hline
		\vspace{-0.2cm}
	\end{tabular}
	\caption{Couplings of $Z$, $Z^{'}$ with electron and top quark in four component spinor mechanism }
	\label{Table2}
	\vspace{-1.2cm}
\end{table} 
 We calculate the diagonal terms  $\left|\mathcal{M}_{\mathcal{D}}\right|^2$ as well as cross terms $\left|\mathcal{M}_{\mathcal{C}}\right|^2$ using amplitudes given above as follows -
 $$\left|\mathcal{M}_{\mathcal{D}}\right|^2 = \frac{1}{4} \sum_{i=\gamma, Z, Z^{'}}^{} \left|\mathcal{M}_{ii}\right|^2; \;\; \left|\mathcal{M}_{\mathcal{C}}\right|^2 = \sum_{\substack{i,j=\gamma, Z, Z^{'}\\i\neq j}}  \frac{1}{2}\sum \;\text{Re} \left[\mathcal{M}_{i}^\dagger \mathcal{M}_{j}\right]$$  
 We compare the result of diagonal term for $Z^{'}$ as well as the cross terms for $Z^{'}$ boson by comparing the the coefficients corresponding product of momenta, eg. $s^{2} p_2\cdot p_3, s^{2} p_2\cdot p_4, s^{2} p_1\cdot p_4$ and $ s^{2} p_1\cdot p_3  $ in both the equations, and get the following relations between the couplings:
 \begin{eqnarray}
 \hspace{-0.2cm}	8 {a_e^{'}}^2 {b_t^{'}}^2 + 8 {a_t^{'}}^2 {b_e^{'}}^2  = \left(C_{Ae}^{'2} + C_{Ve}^{'2} \right) &~&  \hspace{-0.6cm}\left(C_{At}^{'2} + C_{Vt}^{'2}\right)  - 4 C_{Ae}^{'} C_{At}^{'} C_{Ve}^{'} C_{Vt}^{'}\; \; \label{eq1} \; \; \;\\
 	8 {a_e^{'}}^2 {a_t^{'}}^2 + 8 {b_e^{'}}^2 {b_t^{'}}^2 = \left(C_{Ae}^{'2} + C_{Ve}^{'2}\right) &~&  \hspace{-0.6cm}\left(C_{At}^{'2} + C_{Vt}^{'2}\right) + 4 C_{Ae}^{'} C_{At}^{'} C_{Ve}^{'} C_{Vt}^{'}\; \; \label{eq2}\\
 	4 \left({a_e^{'}}^2 {a_t^{'}}^2+ {b_e^{'}}^2 {b_t^{'}}^2\right) \left(2 M_{Z}^2-s\right)+ 4 s\left({a_e^{'}}^2 {b_t^{'}}^2+  {a_t^{'}}^2 {b_e^{'}}^2\right)   &=& \nonumber \\  M_{Z'}^2 \left(C_{Ae}^{'2} +C_{Ve}^{'2} \right)   \left(C_{At}^{'2}+C_{Vt}^{'2}\right) &~& \hspace{-0.4cm}+4 C_{Ae}^{'} C_{At}^{'} C_{Ve}^{'} C_{Vt}^{'}
 	\left(M_{Z'}^2-s\right) \\
 	C_{Ve}^{'} C_{Vt}^{'} + C_{Ae}^{'} C_{At}^{'} & = & 2 a_e^{'} a_t^{'} + 2 b_e^{'}   b_t^{'}	\nonumber \\
 	C_{Ve}^{'} C_{Vt}^{'} - C_{Ae}^{'} C_{At}^{'} &=& 2 a_t^{'}   b_e^{'} + 2 a_e^{'} b_t^{'} \nonumber \\
 	C_{Ve}^{'} C_{Vt}^{'} &=& b_e^{'} b_t^{'} + a_e^{'} a_t^{'} + a_t^{'}   b_e^{'} + a_e^{'} b_t^{'}
 	\label{eq3}
 \end{eqnarray}
 The solution to above equations  is given by -
 \begin{eqnarray}
 	\hspace{-0.5cm} C_{Ve}^{'} &=& a_{e}^{'} + b_{e}^{'}, \hspace{1cm} C_{Ae}^{'} = a_{e}^{'} - b_{e}^{'},  \hspace{1cm}
  C_{Vt}^{'} = a_{t}^{'} + b_{t}^{'} \hspace{1cm} C_{At}^{'} = a_{t}^{'} - b_{t}^{'} \;\;\;
 	\label{soln}
 	 \vspace{-0.3cm}
 \end{eqnarray}
\section{Conclusion}
\vspace{-0.2cm}
In this article, we have shown the equivalence between the two component spinor and four component spinor expressions of $S$ matrix elements for the process $e^{-} e^{+} \rightarrow t \Bar{t}$ at the tree level, taking into account scattering mediated by three neutral bosons namely standard photon, Z and a NP $Z^{'}$. We work out the relations between vector and axial vector couplings of electron and top quark to $Z^{'}$ in both the mechanisms.

\end{document}